\documentstyle[aps]{revtex}\newcommand{\be}{\begin{equation}}

\newcommand{\ee}{\end{equation}}
\newcommand{\bea}{\begin{eqnarray}}
\newcommand{\eea}{\end{eqnarray}}

\font\mybb=msbm10 at 10pt
\def\bb#1{\hbox{\mybb#1}}
\def\bZ {\bb{Z}}
\def\bR {\bb{R}}
\def\bE {\bb{E}}

\def\bC {\bb{C}}

\begin{document}
\twocolumn[\hsize\textwidth\columnwidth\hsize\csname
@twocolumnfalse\endcsname

\date{May 1999, revised July 1999}
\rightline{DAMTP-1999-68}
\rightline{hep-th/9905196}
\title{A Bogomol'nyi equation for intersecting domain walls}
\author{G.W. Gibbons and {}~P.K. Townsend }
\address{
DAMTP, Univ. of Cambridge, Silver St., Cambridge CB3 9EW,  UK\\
 }
\maketitle
\begin{abstract}
We argue that the Wess-Zumino model with quartic superpotential admits
stable static solutions in which three domain walls intersect at a junction. We
derive an energy bound for such junctions and show that configurations
saturating it preserve 1/4 supersymmetry. 
\end{abstract}
\vskip2pc]


Domain walls arise in many areas of physics. They occur as solutions 
of scalar field theories whenever the potential is such that it has isolated
degenerate minima. There are two circumstances in which this happens naturally.
One is when a discrete symmetry is spontaneously broken; in this case the
degeneracy is due to the symmetry. The other is when the field theory is
supersymmetric; in this case the potential is derived from a 
superpotential, the critical points of which are degenerate minima of the
potential. A simple model illustrating the latter case is the 
3+1 dimensional (or D=4) Wess-Zumino (WZ) model with a superpotential $W(\phi)$
that is a polynomial function of the complex scalar field $\phi$. The static 
domain wall solutions of this theory \cite{AT,cvetic} are stable for
topological reasons but the stability can also be deduced from the fact that
a static domain wall is `supersymmetric', i.e. it partially preserves the
supersymmetry of the vacuum. An advantage of the latter point of view is
that the condition for supersymmetry leads immediately to a first
order `Bogomol'nyi' equation, the solutions of which automatically
solve the second order field equations. 

The possibilities for partial preservation of supersymmetry in the WZ
model can be analysed directly in terms of the N=1 D=4
supertranslation algebra. Allowing for all algebraically 
independent central charges, the matrix of anticommutators of the
spinor charge components $S$ is
\be
\{S,S\} = H + \Gamma^{0i}P_i + {1\over2}\Gamma^{0ij}U_{ij} +
{1\over2}\Gamma^{0ij}\gamma_5 V_{ij}
\ee
where $H$ is the Hamiltonian, $P_i$ the 3-momentum, $U$ and $V$ are two
2-form charges, $(\Gamma^0,\Gamma^i)$ are the $4\times 4$ Dirac matrices
and $\gamma_5=\Gamma^{0123}$. The fraction of supersymmetry preserved by any
configuration carrying these charges is one quarter of the number of
zero-eigenvalue eigenspinors $\zeta$  of the matrix $\{S,S\}$. Supersymmetric
configurations other than the vacuum will preserve either 1/2 or 1/4 of the
supersymmetry. A domain wall in the 1-3 plane, for example, has
$(U_{13},V_{13})=H(\cos\alpha,\sin\alpha)$ for some angle
$\alpha$ \cite{AT,WO}; the corresponding spinors $\zeta$ are eigenspinors 
of $\Gamma^{013}\exp(\alpha\gamma_5)$ from which it 
follows that the domain wall preserves 1/2
supersymmetry. Now consider a configuration with non-zero $H$, $U_{13}= u$,
$V_{23}=v$, all other charges vanishing; such a configuration preserves 1/4
supersymmetry if $|u|+|v|=H$, with $\zeta$ an eigenspinor 
of both $\Gamma^{013}$
and $\Gamma^{023}\gamma_5$. Such a configuration would naturally be associated
with domain walls in the 1-3 and 2-3 planes intersecting on the 3-axis. 
In this paper we argue that this possibility is realized in the WZ model.

Intersections of domain walls have been extensively studied in the
context of a theory with a single real scalar field $\varphi$ on $\bE^3$
\cite{E,SZ,BR}. Static configurations are presumed to satisfy an
equation of the form
\be\label{motion}
\nabla ^2 \varphi = V^\prime (\varphi)\, ,
\ee
where $\nabla^2$ is the Laplacian on $\bE^3$ and $V(\varphi)$ is 
a real positive function of $\varphi$ with two
adjacent isolated minima at which $V$ vanishes. Let 
these minima be at $\varphi=\pm1$ and let
$(x,y,w)$ be cartesian coordinates for $\bE^3$.   
If one assumes that $\varphi \rightarrow \pm1$ as $x \rightarrow \pm
\infty$, uniformly in $y$ and $w$, then solutions of (\ref{motion}) 
are necessarily planar because {\sl all} \cite{X,Y,V,W,U} such
solutions satisfy the first order ordinary differential equation
\be\label{bogone}
{ d\varphi \over dx }= \sqrt {V}\, .
\ee
The solutions of this equation are the static domain walls which
are stable for topological reasons. In the context of a D=3
supersymmetric model they are also supersymmetric, for reasons
explained at the conclusion of this article

Now consider the possibility of static intersecting domain wall
solutions of (\ref{motion}). An existence proof
has been given \cite{E} showing that (\ref{motion}) admits a solution
representing two orthogonal domain walls. The solution has Dirichlet type
boundary data : $\phi=0$ on the planes $x=0$ and $y=0$
and $\phi \rightarrow \pm 1$ as $|{\bf x}| \rightarrow \infty$
within the first quadrant. Given that  the solution exists
in the first quadrant, it may be obtained in the remaining quadrants 
by reflection. It seems clear, although we are unaware of formal proofs,
that there should also exist solutions for which
$2n$ domain walls intersect, adjacent walls making an angle $\pi/n$.
However all these intersecting solutions are expected to be unstable; it is
certainly the case that they cannot be supersymmetric. 
We shall return to this point later. 

Domain walls with two or more scalar fields have been
investigated in \cite{SZ,BR,BBB}. In \cite{SZ,BR}, three-phase
boundaries were shown to minimise the energy and to correspond, 
in the thin-wall limit, to a `Y-intersection' (with 120 degree
angles). The WZ model is a special case of models of this type.
The energetics of domain wall intersections in the WZ model
was investigated  in \cite{AT} (see also \cite{trodden}). The 
possibilities depend on the form of
the superpotential $W$. If it is cubic then there are two possible domains and
only one type of domain wall separating them. Intersections of two such walls
cannot be more than marginally stable. Stable intersections can
occur only if the superpotential is at least quartic. A quartic 
superpotential can therefore model a tri-stable medium with three possible 
stable domains and three types of domain wall. The 1+1 dimensional 
analysis of \cite{AT} indicates that triple intersections of the three
walls should be stable for some range of parameters,
but static solutions representing such intersections are intrinsically 2+1
dimensional (given that we ignore dependence on the coordinate of the string
intersection), so they cannot be found from the truncation to 1+1
dimensions. However, they should be minimum energy solutions of the 
reduction of the WZ model to 2+1 dimensions. The energy density 
of static configurations in this reduced 2+1 
dimensional theory is
\be\label{ham} 
{\cal H} = {1\over 4} \nabla \phi \cdot \nabla \bar\phi  + |W'(\phi)|^2
\ee
where $\nabla =(\partial_x,\partial_y)$ with $(x,y)$ being 
cartesian coordinates for the two-dimensional space. 

Let $z=x+iy$. The above expression for the energy density 
can then be rewritten as
\be\label{dens}
{\cal H} = \left|{\partial\phi\over\partial z} - 
e^{i\alpha}{\overline {W'}}\right|^2
+ 2{\rm Re}\left( e^{-i\alpha}{\partial W\over \partial z}\right)
+ {1\over2} J(z,\bar z)
\ee
where $\alpha$ is an arbitrary phase, and
\be
J(z,\bar z) =
\left({\partial\phi\over\partial\bar z}\,
{\partial\bar\phi\over\partial z} - {\partial\phi\over\partial z}\,
{\partial\bar \phi\over\partial\bar z}\right)\, .
\ee
We now observe that 
\be
Q \equiv {1\over2}\int\! dxdy \, J(z,\bar z) = \int\!\Omega\, ,
\ee
where $\Omega$ is the 2-form on 2-space induced by the closed
2-form $(i/4)d\bar\phi\wedge d\phi$ on the target space 
(assumed here to be the complex plane).
Since $\Omega$ is real and closed, $Q$ is a real 
topological charge. We may assume
without loss of generality that it is non-negative. Integration over
space then yields the following expression for the energy
\be
E = \int\! dxdy \, \left|{\partial\phi\over\partial z} - 
e^{i\alpha}{\overline {W'}}\right|^2 +  
{\rm Re}\left[e^{-i\alpha}T\right] + Q\, ,
\ee
where $T$ is the complex boundary term
\be
T= 2\int\! dxdy\, {\partial W\over \partial z}\, .
\ee
We thereby deduce the Bogomol'nyi-type bound
\be\label{bound}
E \ge Q +  |T|\, , 
\ee
which is saturated by solutions of the `Bogomol'nyi' equation
\be\label{bog}
{\partial\phi\over\partial z} = e^{i\alpha}{\overline {W'}}\, .
\ee

Before considering what solutions this equation may have, we shall first show
that generic solutions preserve 1/4 supersymmetry. The fields of the WZ model
reduced to 2+1 dimensions comprise a complex scalar $\phi$ and a complex
$SL(2;\bR)$ spinor field
$\psi^\alpha$; we use an $SL(2;\bR)$ notation in which
\be
\partial_{\alpha\beta} = \delta_{\alpha\beta}\partial_t + 
(\sigma_1)_{\alpha\beta} \partial_x +
(\sigma_3)_{\alpha\beta}\partial_y
\ee
and $\psi_\alpha = \psi^\beta\varepsilon_{\beta\alpha}$. Similarly,
$\partial^{\alpha\beta} =
\varepsilon^{\alpha\gamma}\varepsilon^{\beta\delta}\partial_{\beta\gamma}$.
The Lagrangian density is
\bea\label{lag}
{\cal L} &=& {1\over 8}\partial^{\alpha\beta}\phi\, 
\partial_{\alpha\beta}\bar\phi
+ {i\over2}\bar\psi^\alpha\partial_{\alpha\beta}\psi^\beta \nonumber\\
&& +\, {i\over2}\left[ W'' \psi^\alpha\psi_\alpha + {\overline {W''}}
\bar\psi^\alpha\bar\psi_\alpha \right] - |W'|^2\, .
\eea
Note that the corresponding bosonic hamiltonian density is precisely
(\ref{ham}). The action is invariant, up a surface term, under the
infinitesimal supersymmetry transformations 
\bea\label{tran}
\delta\phi &=& 2i\epsilon_\alpha \psi^\alpha \nonumber\\
\delta \psi^\alpha &=& - \partial^{\alpha\beta}\phi\, \bar\epsilon_\beta  
-2{\overline {W'}}
\varepsilon^{\alpha\beta}\epsilon_\beta\, ,
\eea
and their complex conjugates (we adopt the convention that bilinears of real
spinors are pure imaginary). 

We see from (\ref{tran}) that purely bosonic configurations are supersymmetric
provided that the equation
\be\label{D3bog}
\partial^{\alpha\beta}\phi\, \bar\epsilon_\beta  + 2{\overline {W'}}
\varepsilon^{\alpha\beta}\epsilon_\beta =0
\ee
admits a solution for some constant complex spinor $\epsilon$. For a 
time-independent
complex field $\phi$ this equation is equivalent to
\be
(1 - \sigma_2) \bar\epsilon\, (\bar\partial\phi) +
(1 + \sigma_2) \bar\epsilon\, (\partial\phi) = 
2{\overline { W'}}\, \sigma_3\epsilon
\ee
where $\partial \equiv \partial/\partial z$ and $\bar\partial \equiv
\partial/\partial \bar z$.  For a field $\phi$ satisfying (\ref{bog}) we deduce
that $\epsilon$ satisfies
\be\label{susyconst}
\sigma_2\bar\epsilon= \bar\epsilon\, , \qquad 
\sigma_3\bar\epsilon = e^{-i\alpha}\epsilon\, .
\ee
These constraints preserve just one of the four supersymmetries. Solutions of
(\ref{bog}) are therefore 1/4 supersymmetric. 

The supersymmetry Noether charge of the above model is the complex
$SL(2;\bR)$ spinor
\be
S = {1\over2}\int\! dxdy \left\{\left[ \dot\phi -
\sigma_1\partial_x \phi -\sigma_3\partial_y \phi\right]\bar\psi
- 2i\sigma_2W'\,\psi \right\} \, .
\ee
We can now use the canonical anticommutation relations of the fermion fields to
compute the anticommutators. After restricting to static bosonic fields one
finds that $\{S,\bar S\}= H-\sigma_2 Q$. Thus the junction 
charge $Q$ appears as
a central charge in the supertranslation algebra. The charge $T$ appears
in the $\{S,S\}$ anticommutator and the positivity of the complete matrix of
supercharges implies the bound (\ref{bound}). Of interest here is how the
junction charge $Q$ appears in the D=4 supersymmetry algebra from which we
started. It appears in the same way as would the $P_3$ component of the
momentum and is asociated with the constraint $\Gamma^{03}\zeta=-\zeta$.
This constraint is equivalent to $\Gamma^{023}\gamma_5\zeta=\zeta$ on the $+1$
eigenspace of $\Gamma^{013}$ so, indirectly, we have found a field theory
realization of the 1/4 supersymmetric charge configurations that we
earlier deduced from the N=1 D=4 supersymmetry algebra alone. 

We now return to the `Bogomol'nyi' equation (\ref{bog}). 
If $\phi$ is restricted to be a function only of $x$ then this
equation reduces to the one studied in \cite{AT}, which admits domain wall
solutions parallel to the y-axis. Each domain wall is associated 
with a complex topological charge of magnitude $|\int dx\,\partial_x W|$ and
phase $\alpha$. The question of stability of domain wall junctions 
was adressed 
in \cite{AT} by asking whether two domain walls parallel to the 
y-axis, at least locally, can fuse to form a third
domain wall of lower energy. It was found that this is possible only if their
phases differ; otherwise, stability is marginal. Given that the
energetics allows the formation of an intersection, we would like to
find the static intersecting domain wall solution to which the system
relaxes. Such solutions must depend on {\sl both} 
$x$ and $y$ (equivalently, on both $z$ and $\bar z$), and hence are
much harder to find.

To simplify our task, we shall consider the simple quartic superpotential, 
\be\label{pot}
W(\phi)= -{1\over 4}\phi^4 + \phi\, .
\ee
This has three critical points, at $\phi=1,\omega, \omega ^2$, and a $\bZ_3$
symmetry permuting them. There are therefore three possible domains and three
types of domain wall separating them. The Bogomol'nyi equation
corresponding to this superpotential is
\be\label{tricky}
{\partial \phi \over \partial z} = 1 - \bar \phi^3 \, .
\ee
We have set the phase $\alpha=1$ since it can now be removed by a
redefinition of $z$. This equation is invariant under the $\bZ_3$ action : 
$(z, \phi) \rightarrow (\omega z, \omega \phi)$, so we are led to
seek a $\bZ_3$ invariant solution such that $\phi \rightarrow 1$ as one goes to
infinity inside the sector  $-{\pi \over 6} < \arg z < {\pi \over 6}$,
subject to the condition that $\arg \phi= \arg z$ on the boundary. 
By symmetry $\phi$ must vanish at the origin and so 
$\phi \approx z$ for small $z$.
Given that a stable static triple intersection exists, there should also exist 
meta-stable networks of domain walls \cite{note}. For example, 
one may imagine a static lattice consisting of hexagonal domains, rather 
like graphite. The vertices form triple
intersections and  one may consistently label the hexagons of the array with
$1,\omega, \omega ^2$, in such a way that no two domains which touch along a
common edge carry the same label. The evolution of networks of domain
walls has been studied numerically in \cite{RPS}. We believe that it
would be fruitful to study the WZ model in this context.   

It is well known that topological defects such as strings and domain walls
admit wavelike excitations travelling along them at the speed of light.
The domain wall junctions considered here are no exception.
One easily checks that the D=4 WZ equations are satisfied
if $\phi(z)$ solves  our Bogomol`nyi equation (\ref{bog}) but
is also allowed to have arbitrary dependence upon {\sl either} $t-w$ {\sl or}
$t+w$, where $w$ is the third space coordinate on which we reduced to get the
(2+1) dimensional model.  However, {\sl only one choice preserves
supersymmetry}. To see this we note that the $SL(2;\bC)$-invariant 
condition for preservation of supersymmetry in the unreduced D=4 WZ model is
\be
\partial^{\alpha\dot\beta}\phi\, \bar\epsilon_{\dot\beta} + 2{\overline {W'}}
\varepsilon^{\alpha\beta}\epsilon_\beta =0\, . 
\ee
Given that the reduced D=3 equation (\ref{D3bog}) is satisfied, and that
the spinor $\epsilon$ satisfies (\ref{susyconst}), we then deduce that
\be
\partial_+ \phi\,  \bar \epsilon =0
\ee
where $\partial_+= \partial_t\pm \partial_w$, the sign depending on the choice
of conventions. Thus, we again have 1/4 supersymmetry if $\partial_+\phi=0$ but
no supersymmetry if $\partial_-\phi=0$. This result is not unexpected because
we saw earlier that the junction charge $Q$ appears in the D=4 supersymmetry
algebra in the same way as does $P_3$.

Note that since an individual domain wall preserves 1/2 supersymmetry its low
energy dynamics must be described by a (2+1)-dimensional supersymmetric field
theory with two supersymmetries (corresponding to N=1). The two components of
the spinor field of this effective theory are the coefficients of two
Nambu-Goldstone fermions associated with the broken supersymmetries.
The domain wall junction preserves only 1/4 supersymmetry, so there must be a
total of three Nambu-Goldstone fermions localised on the intersecting domain
wall configuration as a whole. Only two of these are free to propagate within 
the walls, so the third Nambu-Goldstone fermion must be localised on the string
junction. This can also be seen by viewing the junction as a 1/2-supersymmetric
defect on a given wall. The fact that half of the wall's supersymmetry is
preserved means that the junction's low energy dynamics is
described by a (1,0)-supersymmetric (1+1) dimensional field theory.
This theory is chiral with one fermion that is either left-moving or
right-moving; let us declare it to be left-moving. This fermion is the
Nambu-Goldstone fermion associated with the fact that the junction also breaks
half the wall's supersymmetry. Its bosonic partner under (1,0) supersymmetry
must also be left-moving. It follows that right-moving waves are
supersymmetric whereas left-moving ones are not, precisely as we deduced above
from other considerations. 

Now that we have a good understanding of the pattern of supersymmetry breaking
in the WZ model we return to the simpler model discussed earlier with one real
scalar field. This model has an N=1 supersymmetrization in (2+1) dimensions,
with $V= 4(W')^2$, obtained by restricting all quantities in the N=2 model
discussed above to be real. Taking $\partial_y\phi=0$ we then find that 
solutions of (\ref{bogone}) are supersymmetric,
with the real 2-component spinor $\epsilon$ an eigenspinor of 
$\sigma_3$. This might seem paradoxical in view of
the fact that the N=1 D=3 supertranslation algebra admits no
central charges, of either scalar or vector type, that are 
{\sl algebraically} independent of the 3-momentum. The resolution 
is that the anticommutator of supersymmetry charges $S_\alpha$ is 
\bea
\{S_\alpha,S_\beta\} & =& \delta_{\alpha\beta}H \nonumber\\
&+&
(\sigma_1)_{\alpha\beta}\left(P_x+ T_y\right) +
(\sigma_3)_{\alpha\beta}\left(P_y- T_x\right)
\eea
where $H$ is the Hamiltonian, ${\bf P}$ is the field 2-momentum and
${\bf T} = \int d^2x \nabla W$ is a 2-vector topological charge (the
corrresponding algebra of currents was discussed in \cite{jerome}). 
For static solutions ${\bf P}$ vanishes, while $\partial_y W$ vanishes for
solutions with $\partial_y\phi=0$. For such solutions we have
\be
\{S,S\} = H + \sigma_3T_x\, .
\ee
It follows that $H\ge |T_x|$. Field configurations that saturate this bound
preserve 1/2 the supersymmetry and are associated with eigenspinors of 
$\sigma_3$, as claimed.  
An intersecting domain wall solution in this N=1 (2+1)-dimensional model 
cannot satisfy (\ref{bogone}) (because its only static solutions are the
planar domain walls) and this means that it cannot be supersymmetric. In 
contrast to the model with a complex scalar, one cannot use
supersymmetry to argue for the stability of domain wall junctions 
in a model with only one real scalar field. 
\vskip 0.3cm

\noindent
{\bf Acknowledgements}: GWG thanks Martin Barlow and Alberto Farina
for informing him of their, and related, work on domain walls. We are
also grateful for conversations with Paul Shellard and 
David Stuart.



\begin{thebibliography}{99}

\bibitem{AT} E R C Abraham and P K Townsend,
{\sl Intersecting Extended Objects in Supersymmetric Field Theories}
{\it Nucl Phys }{\bf B351} (1991) 313-332. 

\bibitem{cvetic} M. Cvetic, F. Quevedo and S.J. Rey, {\sl String
domain walls and target space modular invariance},
Phys. Rev. Lett. {\bf 67} (1991) 1836-1839.

\bibitem{WO}
The association of a domain wall with a 2-form charge in the 
supersymmetry algebra is expected from the model-independent analysis 
of de Azcarraga et al. (Phys. Rev. Lett. {\bf 63} (1989) 129) based 
on the supermembrane action. It is a generalization of the result of
Witten and Olive ({\it Phys. Lett.} {\bf 78B} (1978) 97) for scalar charges. 
More recently, the 2-form charge has been rediscovered, and its
importance emphasized, by Dvali and Shifman 
(Phys. Lett. {\bf 396B} (1997) 64) in the context of SQCD. 

\bibitem{E} H Dang, P C Fife and L A Peletier, {\sl Saddle solutions
of the bistable diffusion equation}, {\it Z Angew Math Phys} {\bf 43}
(1992) 984-998.

\bibitem{SZ} 
P Sternberg and W P Zeimer, {\sl Local minisers of a three-phase 
problem with triple junctions}, 
{\it Proc Roy Soc Edinburgh}  {\bf A 124} (1994)
1059-1073

\bibitem{BR} 
L Bronsard and F Reitich, {\sl On three-phase boundary motion and 
the singular limit of  vector-valued Ginzburg-Landau equation}, {\it
Arch Rational Mech Anal} {\bf 124} (1993) 355-379

\bibitem{X} M T Barlow, R F Bass and C Gui, 
{\sl The Liouville property and a conjecture of de Giorgi}, {preprint}

\bibitem{Y} A Farina, {\sl Symmetry for semilinear elliptic 
equations in $\bR^N$ and related conjectures}, preprint

\bibitem{V} N Ghoossoub and C Gui, {\sl On a conjecture of de Giorgi 
and some related problems},  
{\it Math Ann} {\bf 311} (1998) 481-491.

\bibitem{W} H Berestycki, F Hamel and R Monneau, {\sl One-dimensional 
symmetry of bounded entire solutions of some elliptic equations}, preprint

\bibitem{U} G Carbou, {\sl Unicit\'e et minimalit\'e des solutions 
d'une \'equation de Ginzburg-Landau}, {\it Ann Inst Henri Poincar\'e}
{\bf 12} (1995) 305-318.

\bibitem{BBB} 
D Bazeia, H Boschi-Filho and F A Brito, {\sl Domain defects in 
systems of two real scalar fields}, {\it JHEP} 9811(1998) 026,  
hep-th/9811084.

\bibitem{trodden}
S.M. Carroll and M. Trodden, {\sl Dirichlet Topological Defects},
Phys. Rev. {\bf D57} (1998) 5189-5194.

\bibitem{note}
Shortly after submission of this paper to the
archives, one by Carroll, Hellerman and Trodden appeared (hep-th/9905217)
having a significant overlap with the work reported here. 
These authors have also pointed out to us that networks of domain walls 
connected by 1/4 BBS junctions are not 
themselves BPS and can decay by a tunelling process. We agree. We owe
the following argument to P. Saffin, who has confirmed the
meta-stability  numerically. 
For a given orientation, each `Y' junction of the model we discuss is 
actually one of six possible types, corrresponding to the 
action of $\bZ_3$ on a a given `Y' junction and its anti-junction, 
for which $\partial \phi/ \partial z$ is replaced by
$\partial\phi/\partial \bar z$ in (\ref{tricky}). In a hexagonal array, each of
the six junction types occurs once. Since they are each approximate
solutions of different (albeit isomorphic) Bogomol'nyi equations, the
network cannot be BPS.

\bibitem{RPS}
B.S. Ryden, W.H. Press and D.N. Spergel, {\sl The evolution of
networks of domain walls and cosmic strings}, {\it Ap J} {\bf 357}
(1990) 293-300. 

\bibitem{jerome}
J.P. Gauntlett, {\sl Current algebra of the D=3 superstring and
partial breaking of global supersymmetry}, Phys. Lett. {\bf 228B}
(1989) 188-192. 

\end{thebibliography}
\end{document}